\documentclass[conference, letterpaper]{IEEEtran}

\ifCLASSINFOpdf
   \usepackage[pdftex]{graphicx}
\else
\fi

\usepackage[cmex10]{amsmath}
\usepackage{color}
\usepackage{amsmath,amssymb,amsfonts}
\usepackage{algorithmic}
\usepackage{graphicx}
\usepackage{textcomp}
\usepackage{xcolor}
\usepackage{fontawesome}
\usepackage{url}
\usepackage{fancyhdr}
\usepackage[caption=false,font=footnotesize]{subfig}

\renewcommand{\thispagestyle}[2]{}

\fancypagestyle{plain}{
        \fancyhead{}
        \fancyhead[C]{first page center header}
        \fancyfoot{}
        \fancyfoot[C]{first page center footer}
}
\begin{document}

\title{NOISE REDUCTION IN MEDICAL IMAGES\\}

\author{\IEEEauthorblockN{Florez-Aroni Sussana M. }
\IEEEauthorblockA{Faculty of Statistic and Computer Engineering,\\ Universidad Nacional del Altiplano de Puno, P.O. Box 291\\ Puno - Peru.\\
Email: sfloreza@est.unap.edu.pe }
\and
\IEEEauthorblockN{Hancco-Condori Mijail A. }
\IEEEauthorblockA{Faculty of Statistic and Computer Engineering,\\ Universidad Nacional del Altiplano de Puno, P.O. Box 291\\ Puno - Peru.\\
Email: mijax45@gmail.com }
\and
\IEEEauthorblockN{Torres-Cruz Fred}
\IEEEauthorblockA{Faculty of Statistic and Computer Engineering,\\ Universidad Nacional del Altiplano de Puno, P.O. Box 291\\ Puno - Peru.\\
Email: ftorres@unap.edu.pe}}

\maketitle

\begin{abstract}
Objectives: Analyze the types of studies and algorithms that are most applied, Identify the anatomical regions treated. Determine the application of parallel techniques used in studies carried out between 2010 and 2022 in research on noise reduction in medical images.
Methodology: A systematic review of the literature on noise reduction in medical images in the last 12 years was carried out. The observation technique was applied to extract the information and the indicators (type of study, treated anatomical region, algorithm and/or method and the application of parallel computing) were recorded in a data sheet.
Results: Most of the studies have been developed in anatomical regions such as: Brain, Bones, Heart, Breast, Lung and Visual system. In the articles investigated, 14 are applied through parallel computing.
Concluti\'on:  Noise reduction in medical images can contribute to better quality images and thus make a more accurate and effective diagnosis.
\end{abstract}


\IEEEpeerreviewmaketitle
\section{Introduction}
Medical imaging such as MRI comprises one of the most prominent imaging categories. Doctors and other specialists use magnetic resonance imaging to diagnose brain diseases. The categorization of these images using computer systems can be a significant contribution to this field that would be useful for both clinicians and patients. Such a system can help physicians improve the speed and accuracy of their diagnoses.\cite{TANDEL2020103804}

Digital image processing consists of algorithmic processes that transform an image, the main task of image processing is noise reduction in medical images. \cite{cadena2017}. Noise causes degradation in image quality, which can affect the accuracy of medical diagnoses\cite{reduccion2}.

Noise in medical images is commonly caused by the malfunction of the sensor and other hardware in the process of forming, storing or transmitting images, this type of noise affects some individual pixels, changing their original values.\cite{arnal20} 

The noise appears as a granular pattern due to coherent constructive and destructive interference from scattered echoes reflected by microscopic scattering from tissue. The resulting granular pattern reduces the contrast resolution and detail obtainable within the tissue and creates a negative effect on various image interpretation tasks such as feature extraction, segmentation, registration, savannah, recognition, volume rendering, and 3D visualization. Therefore, the removal of spots through digital image processing should improve the image quality and diagnostic potential of medical images.\cite{arnal20}

Various noise reduction methods and algorithms have been applied to improve the quality of medical images, such as:
Parallel computing proposes a new hybrid algorithm for the reduction of speckle noise in medical images. The method efficiently combines the advantages of various denoising filters using local and non-local information. A comparative study is carried out using quantitative and qualitative measures and showing the competitiveness of the proposed method. Both methods are parallelized using OpenMP and a hybrid combination of MPI and OpenMP.\cite{arnal20}
In this systematic review, 37 scientific articles are analyzed for the reduction of noise in medical images. Therefore, the objectives of the study are: a) Analyze the Algorithms applied, b) identify the anatomical regions treated and c) determine the type of study used in noise reduction studies in medical images carried out between 2010 and 2022 as described. Observe in our data sheet.

\section{METHODOLOGY}
\subsection{Type of study} 
A systematic review of the literature on noise reduction in medical images in the last 12 years was carried out. The search for information was considered from the years 2010-2022.
The studies are published in English and Spanish and have been carried out all over the world. The keywords used were: noise, digital processing, medical images, parallel computing.

\subsection{Techniques and instruments}
The observation technique was used to systematize the original articles. A data sheet was prepared to record the information. The indicators used were author, year, algorithms, methodology, parallel computing, anatomical region treated.

\subsection{Literature search procedure}
Information was collected from the repository database:
\hfill

Scopus

\url{https://www.scopus.com}
\hfill

Academic google

\url{https://scholar.google.es/schhp?hl=es}
\hfill

Science Direct

\url{www.sciencedirect.com}
\hfill
\begin{center}
   {\includegraphics[width=0.4\textwidth]{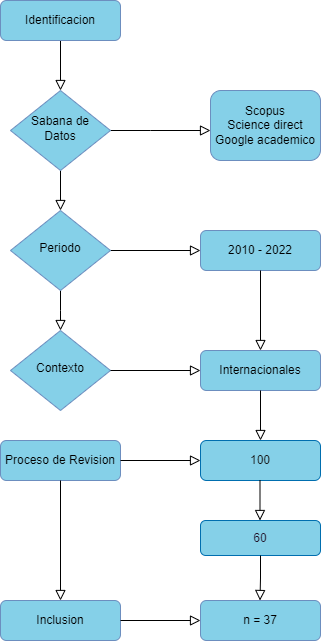}\par} 
\end{center}
\hfill

Prism Flowchart: Information search process for the systematization of original articles 2010-2022.
\hfill

\section{ALGORITHM}
The process of organizing the data sheet of the articles was entered into an Excel template. From this, the information was systematized through qualitative and/or quantitative analysis.

\subsection{Neural network}
AlexNet and VGGNet are deep convulsive neural network (CNN) models that are frequently used in machine learning and computer vision. AlexNet and VGGNet models have both been used successfully in a wide range of computer vision tasks, including image classification and object detection.
\subsection{Parallel ART algorithm}
The Algebraic Reconstruction Technique (ART) algorithm is a technique used in computed tomography image reconstruction (CT). Computed tomography (CT) is a medical imaging technique that uses X-rays to create detailed images of internal organs and structures. It is used to reconstruct CT images from X-ray data collected during scanning. The algorithm reconstructs the image from this data using algebraic and optimization techniques, yielding a detailed and accurate image of the area of the body being scanned.
\subsection{Wiever Filter Method}
It is a form of image restoration filter used to remove noise from an image and improve its visual quality. To recreate the image more accurately, the Wiener filter uses information about the noise distribution and intensity distribution of the original image.
\subsection{The FISTA method}
(Fast Iterative Shrinkage-Thresholding Algorithm). It is an optimization algorithm used to solve constraint minimization problems. It has been used successfully in a wide variety of applications, including CT image reconstruction and signal processing. It uses iteration and thresholding techniques to quickly and efficiently minimize the loss function.

\section{RESULT}
The studies carried out from 2010 to 2022 are described in Table 1. The 30 studies have developed studies for noise reduction used in medical images in a specific anatomical region. In general, most studies have developed treatments in anatomical areas such as: brain, breast, vision, organs, immune system and in general. Note that 16 studies have conducted their research using parallel computing. The other 21 studies have investigated cases about image processing.
\begin{center}
        {\includegraphics[width=0.5\textwidth]{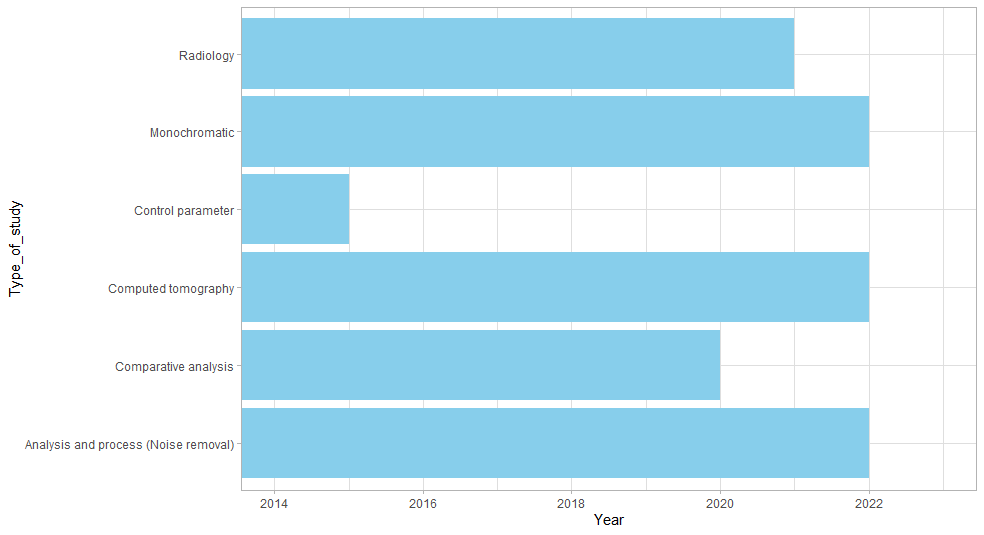}\par}
\end{center}
\hfill

Graphic 1: Type of study from 2010-2022.
\hfill

In graph 1, the 6 categories of the types of studies were obtained, carrying out a deep, objective and logical analysis of the information obtained. It is important to compare the results of this work with previous research, so in this section citations to various references from reliable sources should be observed.

\begin{center}
        {\includegraphics[width=0.5\textwidth]{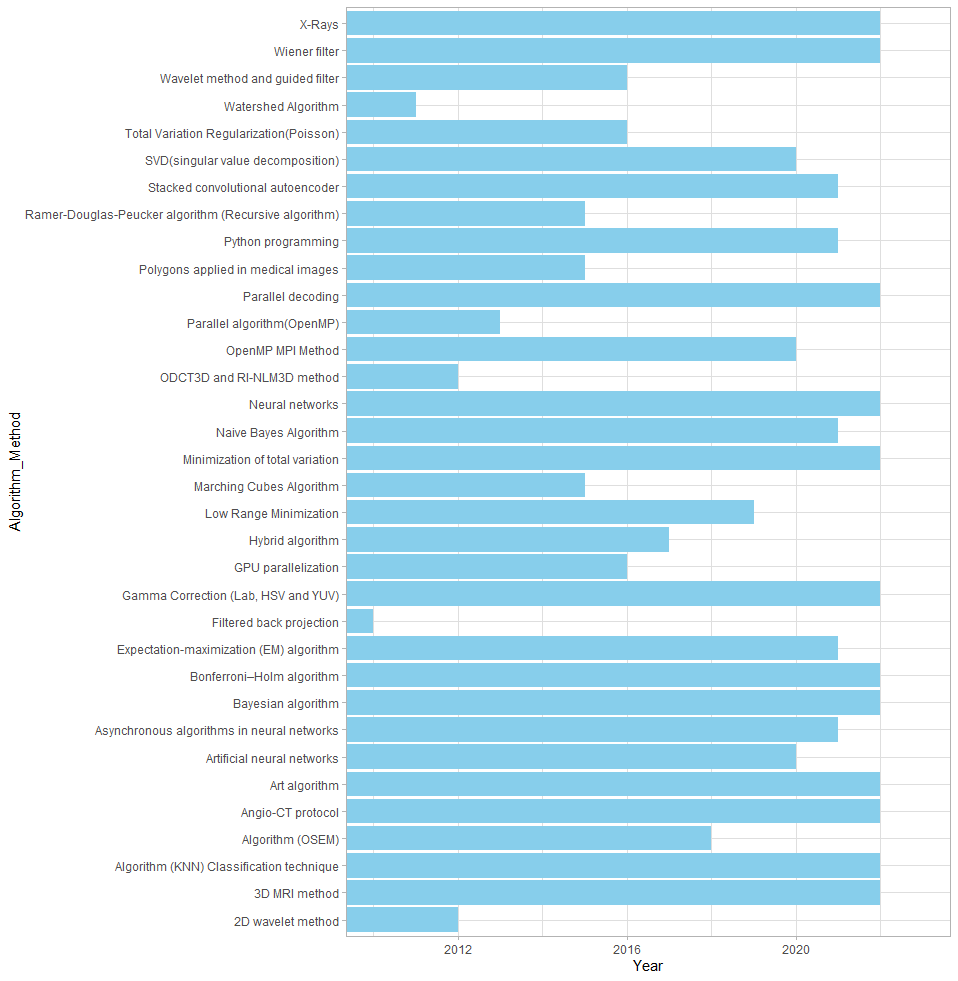}\par}
\end{center}
\hfill

Graphic 2: Classification of Algorithms applied from 2010-2022.
\hfill

\begin{center}
        {\includegraphics[width=0.3\textwidth]{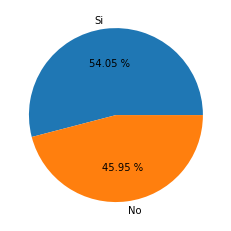}\par}
\end{center}
\hfill
Graphic 3: Algorithms applied in Parallel Computing from 2010-2022.
\hfill

\begin{tabular}{|p{0.2cm}|p{2.5cm}  |p{4.5cm}|}
\hline
N° & Author (Reference) & Contributions  \\ \hline
1 & Ahmad Kazemi, Mohammad Ebrahim Shiri, Amir Sheikhahmadi, Mohamad khodamoradi & Convolutional neural networks are among the most important deep learning methods in which multiple layers are powerfully trained. The proposed method is a deep parallel convolution neural network model consisting of AlexNet and VGGNet networks. \\
\hline

\hline
2 & H. María ShyniE. Chitra. & This article extensively reviews recent deep learning techniques for the diagnosis of COVID-19. The discussed research articles reveal that the Convolutional Neural Network (CNN) is the most popular deep learning algorithm for detecting COVID-19 from medical images.  \\\hline
3 & Andreas Maier, Christophe rSyben, Tobias Lasser, Christian Riess. & This article tries to give a gentle introduction to deep learning in medical image processing, from the theoretical foundations to the applications. We first look at the general reasons for the popularity of deep learning, including several important advances in computing.  \\\hline
4 & Hella Gay D.Avenido Renato V.Crisostomo. & This study implements two types of ART algorithms: sequential and parallel, to speed up the reconstruction of images with a larger number of projections. \\ \hline
5 & Ybargollin-Machado, Perez-Dıaz,Orozco-Morales y Roque-Diaz. &  In this research, 11 modern methods based on total variation are compared for noise reduction in SPECT images. A tomographic section of the Jaszczak mannequin is used, which is processed by all methods.\\ \hline
6 & Volkan Goreke. & He proposes a novel method that removes Poisson noise from medical X-ray images by overcoming the disadvantages.  \\ \hline
7 & Ning Zhang, Long Yu, Dezhi Zhang, Weidong Wu, Shengwei Tian, Xiaojing Kang.& In this article, we propose a new hybrid CNN-transformer structure for the APT-Net medical image segmentation network based on encoder-decoder architecture. \\ \hline
8 & Po Yang, Gordon Clapworthy, Feng Dong, Valeriu Codreanu, David Williams, Baoquan Liu, Jos BTM Roerdink, Zhikun Deng. & It features a GPU-enabled programming model, called GSWO, which can help GPU novices by converting their SWO-based image processing applications from original C/C++ source code to CUDA code in a highly automated manner.  \\ \hline
\end{tabular}

\begin{tabular}{|p{0.2cm}|p{2.5cm}  |p{4.5cm}|}
\hline
9 & Marcos, R.-I., Waldir, S.-H., Neicer, C.-V. & Perform an image classifier and apply a convolutional neuron. After a rigorous training of the neuron to provide an efficient diagnosis, it is shown that by increasing the training the result is more efficient and the error factor is reduced. \\ \hline

\hline
8 & Po Yang, Gordon Clapworthy, Feng Dong, Valeriu Codreanu, David Williams, Baoquan Liu, Jos BTM Roerdink, Zhikun Deng. & It features a GPU-enabled programming model, called GSWO, which can help GPU novices by converting their SWO-based image processing applications from original C/C++ source code to CUDA code in a highly automated manner.  \\ \hline
9 & Marcos, R.-I., Waldir, S.-H., Neicer, C.-V. & Perform an image classifier and apply a convolutional neuron. After a rigorous training of the neuron to provide an efficient diagnosis, it is shown that by increasing the training the result is more efficient and the error factor is reduced. \\ \hline
10 & Sarmiento-Ramos, J. L & It presents an updated review of the main applications of neural networks and deep learning to biomedical engineering in the fields of omics, imaging, brain-machine and human-machine interfaces, and public health management and administration.\\ \hline
11 &Josep Arnall,illa  Mayzel. & A new hybrid algorithm for speckle noise reduction in medical ultrasound imaging is proposed and compared with an existing efficient method. The method efficiently combines the advantages of various denoising filters using local and non-local information.  \\ \hline
12 & MG Sánchez, V. Vidal, J. Bataller, J. Arnal. & The algorithm is based on the peer group concept and uses a fuzzy metric. An optimization study of the use of the CUDA platform to eliminate impulse noise using this algorithm is presented.\\ \hline
\end{tabular}

\hfill
Tabla 2. Selected contributions of the 37 articles
\\ \hfill

\section{DISCUSSION}
In the discussion of results, an in-depth analysis is carried out where studies have shown that in the last 12 years.
The parallel ART algorithms implemented in this study speed up the execution time of image reconstruction compared to the study. Where the implemented parallel ART algorithm had a 42\% increase in its speed compared to its sequential ART algorithm. The acceleration efficiency in the parallel ART algorithms in this study decreases as the number of processors increases. The data dependency of the ART algorithm was not compromised, resulting in an identical reconstructed image for both the parallel algorithm and the sequential algorithm.\cite{AVENIDO2022126}

 The proposed method is a deep parallel convolution neural network model consisting of AlexNet and VGGNet networks. Compared to existing models, the results of the proposed model show that our network has achieved better results. The best result for the database was FIGSHARE, which achieved 99.14\% accuracy in binary class and 98.78\% in multiclass, which was much better than other SVM models. We tested the best available result against the existing database and compared the results with the proposed model, which was the best response of the model for all evaluation criteria. These results suggest that the proposed model can be used as an effective decision support tool for radiologists in medical diagnosis.\cite{KAZEMI2022105775}

The DCRG PPXA method uses a flexible representation of graphical data that allows generalizing the constraint on the projection variable. It shows a new formulation of the VT problem, which can be solved by fast parallel proximal algorithms. The denoising examples show that this method works well on arbitrary graphics, rather than regular grids like SPECT pixels. Consequently, the method is applied to a variety of other inverse problems, including image fusion and mesh filtering.\cite{reduccion2}

In the proposed method it is the Wiener filter that is modified using the FIR filter embedded in the standard Wiener algorithm. The FIR filter design was performed using the ASO optimization algorithm. Local variance optimal and optimal local mean values are calculated using the optimization matrix corresponding to the FIR filter coefficients and transferred to the standard Wiener filter layer as parameter inputs. This method showed superior performance on imaging synthetic and medical X-ray images in terms of PSNR, MSE, SSIM metrics, and image quality metrics such as luminous intensity, contrast ratio, Entropy, and Sharpness. The time consumption of the proposed method is much less.
\cite{GOREKE2023104031}

The FISTA (Fast Iterative Shrinkage/Thresholding Algorithm) method is based on deconvolution methods to reduce noise in images, using computational blur image reconstruction techniques governed by a concise mathematical model. The technique used is within the field of spectral filtering methods. Furthermore, it introduces a necessary regularization or filtering in the reconstructed images.\cite{reduccion2}

\section{Conclution} 
Based on the studies we have read, we have concluded that denoising medical images using deep learning methods has been shown to be more effective than conventional algorithms used to denoise images. This is because the deep neural network is one of the newest and most effective methods to assess medical concussions. The study of this type of articles allows us to become aware of situations that are already beginning to have a great impact on medical health and offers us the opportunity to use more advanced techniques and technologies to improve diagnostic and treatment results.

\end{document}